# Authorized and Unauthorized Practices of Law: The Role of Autonomous Levels of AI Legal Reasoning

**Dr. Lance B. Eliot**

Chief AI Scientist, Techbruim; Fellow, CodeX: Stanford Center for Legal Informatics
Stanford, California, USA

## Abstract

Advances in Artificial Intelligence (AI) and Machine Learning (ML) that are being applied to legal efforts have raised controversial questions about the existent restrictions imposed on the practice-of-law. Generally, the legal field has sought to define Authorized Practices of Law (APL) versus Unauthorized Practices of Law (UPL), though the boundaries are at times amorphous and some contend capricious and self-serving, rather than being devised holistically for the benefit of society all told. A missing ingredient in these arguments is the realization that impending legal profession disruptions due to AI can be more robustly discerned by examining the matter through the lens of a framework utilizing the autonomous levels of AI Legal Reasoning (AILR). This paper explores a newly derived instrumental grid depicting the key characteristics underlying APL and UPL as they apply to the AILR autonomous levels and offers key insights for the furtherance of these crucial practice-of-law debates.

**Keywords:** AI, artificial intelligence, autonomy, autonomous levels, legal reasoning, law, lawyers, authorized practices of law, unauthorized practices of law, judicial system

## 1    Background and Context

Advances in Artificial Intelligence (AI) and Machine Learning (ML) that are being applied to legal efforts have raised controversial questions about the existent restrictions imposed on the practice-of-law [1] [4] [31] [37]. Generally, the legal field has sought to define Authorized Practices of Law (APL) versus Unauthorized Practices of Law (UPL), though the boundaries are at times amorphous and some contend capricious and self-serving, rather than being devised holistically for the benefit of society all told [44] [47].

It is argued herein that a missing ingredient in these debates is the realization that impending legal profession disruptions due to AI can be more robustly discerned by examining the matter through the lens of a framework utilizing the autonomous levels of AI Legal Reasoning (AILR).

Such a grid is presented in this paper and discussed in several respects, including the basis for its formulation, the nature of its utility, and productive opportunities for further extension.

In this paper, five sections are used to cover the topic at hand:

- Section 1: Background and Context
- Section 2: Key Factors of the APL versus UPL
- Section 3: Autonomous Levels of AI Legal Reasoning
- Section 4: APL and UPL Grid Integrating Autonomous Levels of AILR
- Section 5: Additional Considerations and Future Research

In Section 1, an overall background on the matter of APL and UPL is provided. Section 2 then goes further in-depth and identifies what is asserted as key factors underlying APL and UPL. In Section 3, an overview is provided on the autonomous levels of AI Legal Reasoning, crucial to understanding Section 4, which provides a grid that aligns the APL/UPL key factors with the autonomous levels of AI Legal Reasoning. Section 5 is a discussion of additional considerations and also offers suggested avenues for future research on these matters.

## 1.1 Boundaries of APL and UPL

The legal profession has established and continues to maintain that there are Authorized Practices of Law and Unauthorized Practices of Law. Questions regarding the scope and boundaries of APL versus UPL arise with some frequency and especially as technology has advanced, including for example issues surrounding online services such as LegalZoom that provide a claimed capability of producing legal documents by the act of filling in interactive questionnaires [47]. In short, some argue that this kind of service is tantamount to the practice of law and yet is not being performed by an attorney in the act of conducting those services, therefore, it should be considered as overtly unauthorized and deemed ergo unlawful (for legal analyses on these contentions, see for example Shipman [47], Gillers [31], McGinnis [38], and Barton [6]).

Matters such as the LegalZoom controversy remain unsettled and will likely increase in frequency and magnitude as Artificial Intelligence (AI) and especially Machine Learning (ML) are added into these computer-based systems that purport to provide legal services, often referred to as LegalTech [22] [23] [24]. As advances in LegalTech become boosted via AI and ML capabilities, the boundaries of whether those systems are APL or UPL will undoubtedly get further debated and fuel existent disputes over these issues.

One viewpoint on the LegalZoom type of offerings is that as long as such services are only providing static forms and not otherwise seemingly rendering legal advice, they can avoid falling into the unauthorized classification or UPL. As per Barton [6]: "UPL is prohibited in all fifty states. The definition of the 'practice of law' and the levels of enforcement differ from state to state, but at a minimum in no state may a nonlawyer appear in court on behalf of another party. Likewise, nonlawyers may not give 'legal advice.' State bars have long allowed the publication of 'forms books' despite the UPL strictures but have drawn the line at the provision of advice along with forms."

Note that a crucial cornerstone in such an argument entails the rendering or provisioning of legal advice.

Though that might seem like a straightforward restriction, attempts to definitively codify or stipulate exactly what constitutes legal advice has been generally problematic. McClure [37] points out that "state law establishes the parameters of 'the practice of law,' these definitions vary from state to state, but generally, states require bar association admission before either an attorney or a layperson may engage in the practice of law." As such, the shift of attention goes toward whether someone is approved to grant legal advice, as opposed to focusing on what the nature of legal advice itself entails. Similarly, as pointed out further by McClure [37]: "The American Bar Association's Model Rule 5.5 prohibits a person not admitted to the bar association of a particular jurisdiction from practicing law in that specific jurisdiction. A person who is not admitted to the bar association may not represent to the public that he or she may practice law in that jurisdiction."

In essence, avoidance of defining the challenging constructs of "legal advice" is deftly undertaken by sidestepping over into the assertion that only bona fide attorneys can generate or produce legal advice. Thus, the practice of law is seen as that which attorneys do, instead of stating that it is a specified instantiation of legal counsel or legal reasoning involved. Shipman [47] emphasizes the disingenuously distorted logic that this portends in these matters: "It is ironic, given the zealous policing of unauthorized practice of law, that there is not a strong consensus for defining what the practice of law actually is. Comment 2 to Rule 5.5 in the *Model Rules* says that the definition of practice of law is jurisdiction specific and therefore a flexible construct."

This is not to suggest that being able to somehow articulate systematically and with measurable precision the practice-of-law and its constituent elements of legal reasoning can be readily achieved. In fact, it is generally deemed as problematic, resistant to specification, and persistently remains relatively non-standardized. As Shipman [47] aptly explains: "This amorphous standard makes sense given the fact that 'the boundaries of the practice of law are unclear and have been prone to vary over time and geography,' and also because the multifaceted nature of providing legal services makes it difficult to render an exhaustive list of everything the lawyer does in one definition."

In the next subsection, this discussion about these matters will address the various cited bases for why restricting the practice of law is ostensibly justified.

## 1.2 Restrictions on the Practice of Law

The oft-cited rationale for restricting the practice-of-law in terms of only allowing attorneys to undertake such privileges is that this appropriately protects the public and ensures that society is well-served when it comes to justice and the consumption of legal services [45] [51]. The posture taken is that if just anyone was allowed to assert that they were able to practice law, the layperson seeking legal advice might find themselves receiving specious legal advice or worse still outright untoward legal advisement. By keeping the provisioning of legal advice to those certified or authorized to do so, the assumption is that the public will more easily obtain such advice and be less likely to bear the foul fruits of insufficient or improper legal advice. Via a policing function by the legal profession itself, those that have been granted their duly decreed authorization to render legal advice will seemingly be countered and penalized if they violate this instituted capacity [41].

As will be articulated in a moment, besides the rendering of legal advice, there are several additional characteristics opined as essential to the rationale for an overall restriction related to dispensing of legal advice.

Besides seeking to control who can proffer legal advice, the asserted benefits of restricting the practice-of-law encompass other equally vital factors such as creating the venerated lawyer-client relationship and all of its afforded advantages. Per Shipman [47]: "There are many legitimate policy reasons for the restrictions against the unauthorized practice of law. These reasons include 'preserving and strengthening the lawyer-client relationship' and protecting 'the public from being advised and represented in legal matters by unqualified and undisciplined persons over whom the judicial department could exercise slight or no control.' The functioning of the legal system would not be possible without the privileges afforded to and obligations imposed on lawyers when they enter into a formal attorney-client relationship. The formation of an attorney-client relationship subjects a lawyer to 'duties of care, loyalty, confidentiality, and communication, duties' enforceable by the client and through disciplinary sanctions."

Not everyone necessarily views these justifications as being unblemished or quite so pristine.

Some contend the legal profession has put in place rules that amount to a monopolistic effort and ought to be broken apart in an anti-trust manner [6] [31] [32]. Arguments are made that the primary purpose for the APL and UPL is to ensure the economic benefit of attorneys and the law industry and only incidentally exemplifies the nobler claims of seeking to provide a public good. Furthermore, there is the concern that these restrictions are stifling of new innovations, and merely reinforce that law should be practiced as it always has, attempting to keep out any disruption or transformations (this is generally known as the Collingridge [13] innovation conundrum).

Shipman [47] recaps some of these concerns as follows: "Despite the legitimate interests that unauthorized practice of law statutes protect, some critics have rebuked these rules for several reasons. One chief reason is that these rules inhibit innovation in the legal industry. Another major critique is that the bar's purpose in the promulgation of these rules has more to do with protecting lawyers' economic interests than with concerns for the public."

Indeed, the lack of clarity about what the practice of law embodies, and the amorphous notions of legal advice, might be construed as crucial to maintaining the status quo, accordingly stated further by Shipman [47]: "However, overly broad or vague definitions of the practice of law can be detrimental in that they allow lawyers to monopolize certain activities for their own gain and stifle the innovation of affordable alternatives in the world of legal services."

Some legal scholars such as Gillers [31] have examined how lawyers seem to make their own rules in terms of determining what the practice of law is allowed to be, for which might be interpreted as relatively self-serving, and that there ought to be a closer inspection of the rulemaking per se: "What is the responsibility of the profession itself when, through its various institutions and especially bar associations, it asks courts or (less often) lawmakers or agencies to adopt particular rules governing the conduct of lawyers? In other words, my subject is the professional responsibility of the legal profession itself, not the conduct of individual lawyers or the correctness of any particular rule. My purpose is to suggest how the work of devising the rules, not the content of a specific rule, might be improved."

Everything else being equal, the legal profession will presumably be able to continue to keep in place these restrictions, though added pressure will certainly arise due to the advent of AI and ML improvements in existing LegalTech. A key question asked by those in the legal profession is whether the AI advances in LegalTech can be kept at bay and repeatedly exhorted as an illegitimate means of rendering legal advice, despite the potential incremental AI Legal Reasoning (AILR) advances that will emerge [21] [24].

The seemingly easiest way to win that argument is by summarily indicating that the AI LegalTech is not an attorney, and as such, regardless of whether such systems can provide legal advice or not, they cannot be permitted to dispense legal counsel due to the rather axiomatic logic that such AI systems are not, in fact, a lawyer (i.e., overriding any need to examine and nor ascertain whether such AI systems can render proper and legitimate legal advice).

This circular kind of argument might not survive and per McGinnis et al [38] has these potential undermining facets: "The surest way for lawyers to retain the market power of old is to use bar regulation to delay and obstruct the use of machine intelligence. But bar regulation will generally be unavailing. First, lawyers will be able to use many machine-created products to make their own work more cost effective. Thus, using machine inputs can comply with bar regulation, while also creating competitive pressures by lowering costs and reducing the need for the aid of other lawyers. Second, even if unauthorized practice laws in the United States do not change to permit extensive machine intelligence in the production of legal services, those laws will continue to prove ineffective in stemming the emergence of widespread machine lawyering and preserving lawyers' monopoly."

Overall, those points by McGinnis et al [38] suggest that AI LegalTech will potentially be incrementally embraced by attorneys as a vital legal advisement tool, and in so doing will spur AILR advancements more so. Thus, attempts to continue to keep these AI systems in the backend by the law profession overall might momentarily succeed in the nearer term, but those AI systems will be sought for their capabilities and likewise, the vendors will continue to push ahead avidly on advancing them. Presumably, at some future point, the encouragement and enablement on the backend will bring the matter to a head in that eventually those AILR systems might be considered sufficient enough to render legal advice per se, and therefore aim to be unshackled from a backend positioning-only and be repositioned to also encompass the frontend of legal services rendering.

Meanwhile, a second and simultaneous form of pressure might arise by global adoption of AI LegalTech for providing legal advice, doing so in locales that do not have the same restrictions of APL and UPL as does the United States. In this perspective, it is akin to the Genie being let out of the bottle, and some speculate that the prevailing approaches in the U.S. of denying that AI LegalTech can provide independent legal advice will be sorely tested by global adoptions.

In the next section, an in-depth examination of the key characteristics or factors used to shape the APL versus UPL debate is identified and explored.

## 2. Key Factors of the APL versus UPL

Distilling the various characteristics or factors underlying the APL versus UPL debate provides a useful indication of the primary determiners involved. These key factors will be used to then assess how they differ in terms of relevance and impact per a set of autonomous levels entailing AI Legal Reasoning, doing so to illuminate the salient facets of the ongoing dialogue over authorized versus unauthorized practice of law.

There are nine key factors identified, though realize that additional factors can be further gleaned from the myriad of elements utilized in ascertaining APL versus UPL. This core set of nine is evocative of the primary contentions and is sufficient for preparing and providing a grid that can be constructively employed for these discussions. Future research, which is mentioned in the final section of this paper, would be encouraged to consider adding additional key factors, along with subjecting the entire set of factors to an assessment mechanism to potentially rate and appropriately rank their respective significance.

## 2.1 Identified Key Factors

The primary key factors are depicted in a short-form description that is considered suitable for use in a grid and consist of keywords to represent each factor. The key factors consist of:

- Provides Legal Advice
- Asserts Practices Law
- Lawyer-Client Relationship
- Qualified in Law
- Incurs Duty of Care
- Legal Confidentiality
- Enforceable Prof Conduct
- Malpractice Susceptible
- Legal Liability

In the subsections, each key factor will be briefly explained and explored.

## 2.2 Details Underlying Key Factors

For each of the key factors, it is foundational to explain the nature and scope of the factor, doing so to ensure that each can be representative of its focused intent.

### 2.2.1 Provides Legal Advice

The short-form keywords of "Provides Legal Advice" refers to the aspect that ascertaining APL versus UPL involves whether or not there is legal advice that is being proffered. As per the ABA definition [54] of the practice of law and as to the nature of legal advice: "The 'practice of law' is the application of legal principles and judgment with regard to the circumstances or objectives of a person that require the knowledge and skill of a person trained in the law."

This legal advice or practice-of-law arises according to the ABA under these circumstances [54]: "A person is presumed to be practicing law when engaging in any of the following conduct on behalf of another: (1) Giving advice or counsel to persons as to their legal rights or responsibilities or to those of others; (2) Selecting, drafting, or completing legal documents or agreements that affect the legal rights of a person; (3) Representing a person before an adjudicative body, including, but not limited to, preparing or filing documents or conducting discovery; or (4) Negotiating legal rights or responsibilities on behalf of a person."

Presumably, if no legal advice is being rendered, there is no need to analyze whether there is an unauthorized or authorized practice-of-law taking place, simply due to the obvious aspect that there is a lack of legal advice being proffered. On the other hand, if legal advice is involved, potentially any legal advice, even the most infinitesimal, the question then can be dutifully asked about whether this is being done in an authorized versus unauthorized manner.

Whether there is some threshold required as to the significance or magnitude of the legal advice is an open question entailing ongoing research pursuits. For example, if someone makes an offhand remark that would seemingly fit within the scope of the ABA indication of "selecting, drafting, or completing legal documents or agreements that affect the legal rights of the person," does that offhand remark instantaneously invoke that legal advice is being given? Some assert that a kind of reasonableness test needs to be applied to ascertain whether the act has risen to a determinable limit.

### 2.2.2 Asserts Practices Law

The short-form keywords of "Asserts Practices Law" refer to the assertion or communicating that a capability of practicing law exists and that the giving of legal advice can be undertaken by the actor so stating the claimed capacity.

A pertinent ABA provision consists of Rule 7.1 [54]: "A lawyer shall not make a false or misleading communication about the lawyer or the lawyer's services. A communication is false or misleading if it contains a material misrepresentation of fact or law, or omits a fact necessary to make the statement considered as a whole not materially misleading."

Equally pertinent is the ABA Rule 5.5 [54]: "(a) A lawyer shall not practice law in a jurisdiction in violation of the regulation of the legal profession in that jurisdiction, or assist another in doing so. (b) A lawyer who is not admitted to practice in this jurisdiction shall not: (1) except as authorized by these Rules or other law, establish an office or other systematic and continuous presence in this jurisdiction for the practice of law; or (2) hold out to the public or otherwise represent that the lawyer is admitted to practice law in this jurisdiction."

### 2.2.3 Lawyer-Client Relationship

The short-form keywords of "Lawyer-Client Relationship" refers to the aspect that a special relationship is enacted between a lawyer and their client, offering various protections and legal obligations by the lawyer so bounded.

Per the ABA Rule 1.1 [54]: "A lawyer shall provide competent representation to a client. Competent representation requires the legal knowledge, skill, thoroughness and preparation reasonably necessary for the representation," and as stated in Rule 1.3 "A lawyer shall act with reasonable diligence and promptness in representing a client."

### 2.2.4 Qualified in Law

The short-form keywords of "Qualified in Law" refers to the requirement that the legal advisor is appropriately qualified in law.

Per the ABA [54], these are the expected licensing requirements to be an attorney and practice law: "Have a bachelor's degree or its equivalent. Complete three years at an ABA-accredited law school. Pass a state bar examination, which usually lasts for two or three days. The exam tests knowledge in selected areas of law. There are also required tests on professional ethics and responsibility. Pass a character and fitness review. Applicants for law licenses must be approved by a committee that investigates character and background. Take an oath, usually swearing to support the laws and the state and federal constitutions. Receive a license from the highest court in the state, usually the state supreme court."

### 2.2.5 Incurs Duty of Care

The short-form keywords of "Incurs Duty of Care" refers to the need for lawyers to act mindfully when performing their legal acts for clients, and the sufficiency of care is usually evaluated per the prevailing standards of professional competence in law and as applicable to the matter at hand.

As per the ABA indication of a lawyer's responsibilities [54]: "As a representative of clients, a lawyer performs various functions. As advisor, a lawyer provides a client with an informed understanding of the client's legal rights and obligations and explains their practical implications. As advocate, a lawyer zealously asserts the client's position under the rules of the adversary system. As negotiator, a lawyer seeks a result advantageous to the client but consistent with requirements of honest dealings with others. As an evaluator, a lawyer acts by examining a client's legal affairs and reporting about them to the client or to others."

### 2.2.6 Legal Confidentiality

The short-form keywords of "Legal Confidentiality" refers to the confidentiality formed as part of the lawyer-client relationship.

Per ABA Rule 1.6 [54]: "(a) A lawyer shall not reveal information relating to the representation of a client unless the client gives informed consent, the disclosure is impliedly authorized in order to carry out the representation or the disclosure is permitted by paragraph (b)." The aspects of permitted disclosure are stipulated in the Rule 1.6 portion "b" and include various conditions such as confidentiality may be usurped to prevent certain death or substantial bodily harm, to prevent a client from committing a crime or fraud, etc.

### 2.2.7 Enforceable Prof Conduct

The short-form keywords of "Enforceable Prof Conduct" refers to the aspect that there is an expectation of professional conduct by a lawyer in the practice of the law and that this requirement of conduct is enforceable such that if conduct falls below the requisite level then there are adverse consequences that can be imposed upon the attorney so violating the code of conduct.

As per the ABA stipulation [54]: "A lawyer's conduct should conform to the requirements of the law, both in professional service to clients and in the lawyer's business and personal affairs. A lawyer should use the law's procedures only for legitimate purposes and not to harass or intimidate others. A lawyer should demonstrate respect for the legal system and for those who serve it, including judges, other lawyers and public officials. While it is a lawyer's duty, when necessary, to challenge the rectitude of official action, it is also a lawyer's duty to uphold legal process."

## 2.2.8 Malpractice Susceptible

The short-form keywords of "Malpractice Susceptible" refers to a potential failing on the part of the legal advisor to render proper legal advice and to the degree that professional misconduct has occurred and caused harm to another person, making them susceptible to a malpractice claim.

Per the ABA [54]: "Lawyers make mistakes. Sometimes those mistakes have consequences. Ultimately, a viable legal malpractice claim will turn on the facts of the case; but here are three basic things to consider in determining if an attorney's mistake justifies a legal malpractice lawsuit," which encompasses whether the attorney was negligent, whether the mistake caused damage, and whether the damages were significant.

## 2.2.9 Legal Liability

The short-form keywords of "Legal Liability" refers to a wide array of potential liability exposures for attorneys and oftentimes is bucketed into three major facets: (1) disciplinary or violation of legal professional ethics codes, (2) civil claims of liability including malpractice, and (3) criminal claims of liability in the duty of an attorney as an officer of the court and a presumed guardian of the legal system.

In the realm of civil claims of liability, malpractice is singled out in the list of the key factors as shown in the prior subsection. Beyond malpractice, it is customary to consider other acts of liability such as liability for breach of contract, liability for violation of regulatory statutes, and so on. Thus, the value of having a broader category of "Legal Liability" is to ensure that the narrower construing of malpractice would not inadvertently omit or overshadow other forms of legal liability.

## 2.3 Connecting Key Factors With AILR

This section has identified the APL/UPL key factors. The next section describes the autonomous levels of AI Legal Reasoning, providing sufficient context to then align together with the key factors and the LoA AILR in composing an instrumental assessment grid.

## 3.0 Autonomous Levels of AI Legal Reasoning

In this section, a framework for the autonomous levels of AI Legal Reasoning is summarized and is based on the research described in detail in Eliot [25].

These autonomous levels will be portrayed in a grid that aligns with the APL/UPL key factors identified in the prior section of this paper, and thus it is useful to first explain what each of the autonomous levels consists of.

The autonomous levels of AI Legal Reasoning are as follows:

Level 0: No Automation for AI Legal Reasoning
Level 1: Simple Assistance Automation for AI Legal Reasoning
Level 2: Advanced Assistance Automation for AI Legal Reasoning
Level 3: Semi-Autonomous Automation for AI Legal Reasoning
Level 4: Domain Autonomous for AI Legal Reasoning
Level 5: Fully Autonomous for AI Legal Reasoning
Level 6: Superhuman Autonomous for AI Legal Reasoning

See **Figure A-1** for an overview chart showcasing the autonomous levels of AI Legal Reasoning as via columns denoting each of the respective levels.

See **Figure A-2** for an overview chart similar to Figure A-1 which alternatively is indicative of the autonomous levels of AI Legal Reasoning via the rows as depicting the respective levels (this is simply a reformatting of Figure A-1, doing so to aid in illuminating this variant perspective, but does not introduce any new facets or alterations from the contents as already shown in Figure A-1).

## 3.1.1 Level 0: No Automation for AI Legal Reasoning

Level 0 is considered the no automation level. Legal reasoning is carried out via manual methods and principally occurs via paper-based methods.

This level is allowed some leeway in that the use of say a simple handheld calculator or perhaps the use of a fax machine could be allowed or included within this Level 0, though strictly speaking it could be said that any form whatsoever of automation is to be excluded from this level.

### 3.1.2 Level 1: Simple Assistance Automation for AI Legal Reasoning

Level 1 consists of simple assistance automation for AI legal reasoning.

Examples of this category encompassing simple automation would include the use of everyday computer-based word processing, the use of everyday computer-based spreadsheets, the access to online legal documents that are stored and retrieved electronically, and so on.

By-and-large, today's use of computers for legal activities is predominantly within Level 1. It is assumed and expected that over time, the pervasiveness of automation will continue to deepen and widen, and eventually lead to legal activities being supported and within Level 2, rather than Level 1.

### 3.1.3 Level 2: Advanced Assistance Automation for AI Legal Reasoning

Level 2 consists of advanced assistance automation for AI legal reasoning.

Examples of this notion encompassing advanced automation would include the use of query-style Natural Language Processing (NLP), Machine Learning (ML) for case predictions, and so on.

Gradually, over time, it is expected that computer-based systems for legal activities will increasingly make use of advanced automation. Law industry technology that was once at a Level 1 will likely be refined, upgraded, or expanded to include advanced capabilities, and thus be reclassified into Level 2.

### 3.1.4 Level 3: Semi-Autonomous Automation for AI Legal Reasoning

Level 3 consists of semi-autonomous automation for AI legal reasoning.

Examples of this notion encompassing semi-autonomous automation would include the use of Knowledge-Based Systems (KBS) for legal reasoning,

the use of Machine Learning and Deep Learning (ML/DL) for legal reasoning, and so on.

Today, such automation tends to exist in research efforts or prototypes and pilot systems, along with some commercial legal technology that has been infusing these capabilities too.

### 3.1.5 Level 4: Domain Autonomous for AI Legal Reasoning

Level 4 consists of domain autonomous computer-based systems for AI legal reasoning.

This level reuses the conceptual notion of Operational Design Domains (ODDs) as utilized in the autonomous vehicles and self-driving cars levels of autonomy, though in this use case it is being applied to the legal domain [17] [18] [20].

Essentially, this entails any AI legal reasoning capacities that can operate autonomously, entirely so, but that is only able to do so in some limited or constrained legal domain.

### 3.1.6 Level 5: Fully Autonomous for AI Legal Reasoning

Level 5 consists of fully autonomous computer-based systems for AI legal reasoning.

In a sense, Level 5 is the superset of Level 4 in terms of encompassing all possible domains as per however so defined ultimately for Level 4. The only constraint, as it were, consists of the facet that the Level 4 and Level 5 are concerning human intelligence and the capacities thereof. This is an important emphasis due to attempting to distinguish Level 5 from Level 6 (as will be discussed in the next subsection)

It is conceivable that someday there might be a fully autonomous AI legal reasoning capability, one that encompasses all of the law in all foreseeable ways, though this is quite a tall order and remains quite aspirational without a clear cut path of how this might one day be achieved. Nonetheless, it seems to be within the extended realm of possibilities, which is worthwhile to mention in relative terms to Level 6.

### 3.1.7 Level 6: Superhuman Autonomous for AI Legal Reasoning

Level 6 consists of superhuman autonomous computer-based systems for AI legal reasoning.

In a sense, Level 6 is the entirety of Level 5 and adds something beyond that in a manner that is currently ill-defined and perhaps (some would argue) as yet unknowable. The notion is that AI might ultimately exceed human intelligence, rising to become superhuman, and if so, we do not yet have any viable indication of what that superhuman intelligence consists of and nor what kind of thinking it would somehow be able to undertake.

Whether a Level 6 is ever attainable is reliant upon whether superhuman AI is ever attainable, and thus, at this time, this stands as a placeholder for that which might never occur. In any case, having such a placeholder provides a semblance of completeness, doing so without necessarily legitimatizing that superhuman AI is going to be achieved or not. No such claim or dispute is undertaken within this framework.

### 4.0 APL and UPL Grid Integrating Autonomous Levels of AILR

4.1 Grid Indication of Levels of Autonomy (LoA) by Key Factors

In this section, the APL/UPL key factors depicted in Section 2 are aligned into a grid that also contains the autonomous levels of AI Legal Reasoning which were described in Section 3.

**Figure B-1** provides an overview chart depicting the rows as the respective LoA AILR levels and the columns denoting the APL/UPL key factors. A row-by-row explanatory narrative is provided in the subsections below.

**Figure B-2** provides a similar overview chart of Figure B-1 but does so with the rows indicating the APL/UPL key factors and the columns showcasing the APL/UPL key factors. This is simply an alternative perspective of Figure B-1 and does not introduce any new content or alterations from the contents depicted in Figure B-1. A row-by-row explanatory narrative is provided in the subsections below.

### 4.1.1 Level 0: No Automation for AI Legal Reasoning

As indicated in charts B-1 and B-2, Level 0 of the LoA AILR have an "n/a" (meaning "not applicable") for each of the APL/UPL key factors.

This designating of "n/a" seems applicable for Level 0 since there is considered no automation and no AILR autonomy involved at Level 0. As such, there is presumably no opportunity for any potential claim or contention that the automation or autonomy is providing legal advice, and likewise, it is not asserting that it is practicing law, it does not create a lawyer-client relationship, and so on.

Here then is Level 0:

**Level 0**
- Provides Legal Advice: **n/a**
- Asserts Practices Law: **n/a**
- Lawyer-Client Relationship: **n/a**
- Qualified in Law: **n/a**
- Incurs Duty of Care: **n/a**
- Legal Confidentiality: **n/a**
- Enforceable Prof Conduct: **n/a**
- Malpractice Susceptible: **n/a**
- Legal Liability: **n/a**

### 4.1.2 Level 1: Simple Assistance Automation for AI Legal Reasoning

As indicated in charts B-1 and B-2, Level 1 of the LoA AILR is designated as "no" for each of the APL/UPL key factors.

This designating of "no" seems applicable for Level 1 since the automation is considered a simple construct and does not embody any AI autonomous capabilities. Note that though a vendor or developer of such simple legal technology might wish to claim that their system provides legal advice, and for which this is still an open question per the exemplar of LegalZoom discussed in Section 1, for the purposes herein, it is suggested that this is not the case at Level 1, but might be the case at Level 2 (see next subsection).

Here then is Level 1:

**Level 1**
- Provides Legal Advice: **No**
- Asserts Practices Law: **No**
- Lawyer-Client Relationship: **No**
- Qualified in Law: **No**
- Incurs Duty of Care: **No**
- Legal Confidentiality: **No**
- Enforceable Prof Conduct: **No**
- Malpractice Susceptible: **No**
- Legal Liability: **No**

### 4.1.3 Level 2: Advanced Assistance Automation for AI Legal Reasoning

As indicated in charts B-1 and B-2, the Level 2 of the LoA AILR is designated as "no" for the preponderance of the APL/UPL key factors, and has the designation of "maybe" for two of the factors, namely for "Provides Legal Advice" and for "Legal Liability."

This designating of "no" seems applicable for most of Level 2 since the automation does not embody any AI autonomous capabilities.

Despite the lack of AI autonomous capabilities, there is the gray area of whether the automation has entered into the realm of providing legal advice and thus the use of "maybe" as a designator.

Referring to the discussion of Section 1 about LegalZoom as an exemplar, there is still an open question of how far beyond the act of providing a form does it take for the threshold of dispensing legal advice to arise. If there is legal advice being proffered, it would seem logically consequential that an invocation of legal liability could potentially also be raised regarding the legal advice being so offered, and thus the use of "maybe" as a designator for the "Legal Liability" factor.

Here then is Level 2:

**Level 2**
- Provides Legal Advice: **Maybe**
- Asserts Practices Law: **No**
- Lawyer-Client Relationship: **No**
- Qualified in Law: **No**
- Incurs Duty of Care: **No**
- Legal Confidentiality: **No**
- Enforceable Prof Conduct: **No**
- Malpractice Susceptible: **No**
- Legal Liability: **Maybe**

### 4.1.4 Level 3: Semi-Autonomous Automation for AI Legal Reasoning

As indicated in charts B-1 and B-2, the Level 3 of the LoA AILR is designated as a mixture of "no" for many of the APL/UPL key factors, and has the designation of "yes," "minimal," and "likely" for three factors, respectively "Provides Legal Advice," "Qualified in Law," and "Legal Liability."

This designating of "no" seems applicable for much of Level 3 since the automation is only partially embodying AI autonomous capabilities, considered as semi-autonomous. Due to the semi-autonomous nature, it could be argued that systems in Level 3 are providing legal advice, of which presumably do so they need to be qualified in law (at least to some minimal amount), and the provisioning of legal advice would seem to place such systems into the exposure of legal liability for doing so.

Here then is Level 3:

**Level 3**
- Provides Legal Advice: **Yes**
- Asserts Practices Law: **No**
- Lawyer-Client Relationship: **No**
- Qualified in Law: **Minimal**
- Incurs Duty of Care: **No**
- Legal Confidentiality: **No**
- Enforceable Prof Conduct: **No**
- Malpractice Susceptible: **No**
- Legal Liability: **Likely**

### 4.1.5 Level 4: Domain Autonomous for AI Legal Reasoning

As indicated in charts B-1 and B-2, the Level 4 of the LoA AILR is designated as a mixture of "likely" for many of the APL/UPL key factors, and has the designation of "yes" and "partial" for three factors, respectively "Asserts Practices Law," "Lawyer-Client Relationship," "Qualified in Law."

For purposes of nomenclature, the use of the word "partial" and the word "likely" are admittedly somewhat ill-defined and open to interpretation, which is intended for now as to the initial instantiation of this grid. As mentioned in Section 5, it is hoped and anticipated that further research will be undertaken to clarify and more discretely specify these designations.

In Level 4, a significant consideration is that the autonomy of the AILR arises only in selected domain or subdomain strata, thus, there is an inherent restriction or qualification involved. As will be indicated for Level 5 and Level 6, there are no such limits and therefore the use of designators such as "partial" or "likely" are no longer warranted in those levels.

Here then is Level 4:

Level 4
- Provides Legal Advice: **Yes**
- Asserts Practices Law: **Yes**
- Lawyer-Client Relationship: **Partial**
- Qualified in Law: **Partial**
- Incurs Duty of Care: **Likely**
- Legal Confidentiality: **Likely**
- Enforceable Prof Conduct: **Likely**
- Malpractice Susceptible: **Likely**
- Legal Liability: **Likely**

### 4.1.6 Level 5: Fully Autonomous for AI Legal Reasoning

As indicated in charts B-1 and B-2, Level 5 of the LoA AILR are designated as a series of "yes" designations for each of the APL/UPL key factors.

In brief, since the AI Legal Reasoning is considered fully versed at the Level 5, it would seem corresponding that there would be an expectation enveloping the AILR that it ought to comply with the same set of APL/UPL factors as established for human lawyers. There are thorny questions that arise in this indication due to the unclear nature of whether the AILR itself can be held accountable and considered responsible per se, or whether this semblance of assignability is not extendable to AI systems, perhaps being borne instead by others such as those that have developed the AILR or fielded the AILR. These are ongoing and problematic questions, already being earnestly explored in the field of AI and the law, which will undoubtedly continue for quite some time ahead.

Here then is Level 5:

Level 5
- Provides Legal Advice: **Yes**
- Asserts Practices Law: **Yes**
- Lawyer-Client Relationship: **Yes**
- Qualified in Law: **Yes**
- Incurs Duty of Care: **Yes**
- Legal Confidentiality: **Yes**
- Enforceable Prof Conduct: **Yes**
- Malpractice Susceptible: **Yes**
- Legal Liability: **Yes**

### 4.1.7 Level 6: Superhuman Autonomous for AI Legal Reasoning

As indicated in charts B-1 and B-2, the Level 6 of the LoA AILR is designated as a series of "yes" designations for each of the APL/UPL key factors, and three indicating "yes plus," consisting of "Provides Legal Advice," "Asserts Practices Law," and "Qualified in Law."

The basis for providing a "yes plus" designation is that this Level 6 is the as-yet-known superhuman formulation of AI, and presumably, such AI would exceed the human capacity of lawyering. In that light, it seems prudent to suggest that the Level 6 can provide legal advice beyond that of humans, designated as "yes plus," and asserts the practice of law to a "yes plus" accordingly, and surpasses the human boundaries of being qualified for the law too.

Similar to the discussion given about the Level 5 aspect, mentioned in the prior subsection, since the AI Legal Reasoning is considered fully versed at the Level 6 (and even more so versed, at some superhuman capacity), it would seem corresponding that there would be an expectation enveloping the AILR that it ought to comply with the same set of APL/UPL factors as established for human lawyers. As stated about Level 5, there are thorny questions that arise in this indication for Level 6 too, due to the unclear nature of whether the AILR itself can be held accountable and considered responsible per se, or whether this semblance of assignability is not extendable to AI systems, perhaps being borne instead by others such as those that have developed the AILR or fielded the AILR. These are ongoing and problematic questions, already being earnestly explored in the field of AI and the law, which will undoubtedly continue for quite some time ahead.

Here then is Level 6:

### Level 6
- Provides Legal Advice: **Yes Plus**
- Asserts Practices Law: **Yes Plus**
- Lawyer-Client Relationship: **Yes**
- Qualified in Law: **Yes Plus**
- Incurs Duty of Care: **Yes**
- Legal Confidentiality: **Yes**
- Enforceable Prof Conduct: **Yes**
- Malpractice Susceptible: **Yes**
- Legal Liability: **Yes**

## 4.2 Grid Indication of APL/UPL Key Factors by Levels of Autonomy (LoA)

The next subsections showcase the APL/UPL factors as at-a-glance for each factor, listing the designations that have been postulated for each of the LoA AILR levels.

Narrative discussion about these facets has already been covered in the prior Subsection 4.1 and thus it is not necessary to repeat it in this subsection (refer to the prior subsections as needed).

### 4.2.1 APL/UPL "Provides Legal Advice" by LoA

For a narrative discussion about the "Provides Legal Advice" for each of the LoA AILR levels, see the preceding subsections. This list shown here provides the convenience of indication and is also portrayed on charts B-1 and B-2.

#### Provides Legal Advice
- Level 0: **n/a**
- Level 1: **No**
- Level 2: **Maybe**
- Level 3: **Yes**
- Level 4: **Yes**
- Level 5: **Yes**
- Level 6: **Yes Plus**

### 4.2.2 APL/UPL "Asserts Practices Law" by LoA

For a narrative discussion about the "Asserts Practices Law" for each of the LoA AILR levels, see the preceding subsections. This list shown here provides the convenience of indication and is also portrayed on charts B-1 and B-2.

#### Asserts Practices Law
- Level 0: **n/a**
- Level 1: **No**
- Level 2: **No**
- Level 3: **No**
- Level 4: **Yes**
- Level 5: **Yes**
- Level 6: **Yes Plus**

### 4.2.3 APL/UPL "Lawyer-Client Relationship" LoA

For a narrative discussion about the "Lawyer-Client Relationship" for each of the LoA AILR levels, see the preceding subsections. This list shown here provides the convenience of indication and is also portrayed on charts B-1 and B-2.

**Lawyer-Client Relationship**
- Level 0: **n/a**
- Level 1: **No**
- Level 2: **No**
- Level 3: **No**
- Level 4: **Partial**
- Level 5: **Yes**
- Level 6: **Yes**

### 4.2.4 APL/UPL "Qualified in Law" by LoA

For a narrative discussion about the "Qualified in Law" for each of the LoA AILR levels, see the preceding subsections. This list shown here provides the convenience of indication and is also portrayed on charts B-1 and B-2.

**Qualified in Law**
- Level 0: **n/a**
- Level 1: **No**
- Level 2: **No**
- Level 3: **Minimal**
- Level 4: **Partial**
- Level 5: **Yes**
- Level 6: **Yes Plus**

### 4.2.5 APL/UPL "Incurs Duty of Care" by LoA

For a narrative discussion about the "Incurs Duty of Care" for each of the LoA AILR levels, see the preceding subsections. This list shown here provides the convenience of indication and is also portrayed on charts B-1 and B-2.

**Incurs Duty of Care**
- Level 0: **n/a**
- Level 1: **No**
- Level 2: **No**
- Level 3: **No**
- Level 4: **Likely**
- Level 5: **Yes**
- Level 6: **Yes**

### 4.2.6 APL/UPL "Legal Confidentiality" by LoA

For a narrative discussion about the "Legal Confidentiality" for each of the LoA AILR levels, see the preceding subsections. This list shown here provides the convenience of indication and is also portrayed on charts B-1 and B-2.

**Legal Confidentiality**
- Level 0: **n/a**
- Level 1: **No**
- Level 2: **No**
- Level 3: **No**
- Level 4: **Likely**
- Level 5: **Yes**
- Level 6: **Yes**

### 4.2.7 APL/UPL "Enforceable Prof Conduct" LoA

For a narrative discussion about the "Enforceable Prof Conduct" for each of the LoA AILR levels, see the preceding subsections. This list shown here provides the convenience of indication and is also portrayed on charts B-1 and B-2.

**Enforceable Prof Conduct**
- Level 0: **n/a**
- Level 1: **No**
- Level 2: **No**
- Level 3: **No**
- Level 4: **Likely**
- Level 5: **Yes**
- Level 6: **Yes**

### 4.2.8 APL/UPL "Malpractice Susceptible" by LoA

For a narrative discussion about the "Malpractice Susceptible" for each of the LoA AILR levels, see the preceding subsections. This list shown here provides the convenience of indication and is also portrayed on charts B-1 and B-2.

**Malpractice Susceptible**
- Level 0: **n/a**
- Level 1: **No**
- Level 2: **No**
- Level 3: **No**
- Level 4: **Likely**
- Level 5: **Yes**
- Level 6: **Yes**

### 4.2.9 APL/UPL "Legal Liability" by LoA

For a narrative discussion about the "Legal Liability" for each of the LoA AILR levels, see the preceding subsections. This list shown here provides the convenience of indication and is also portrayed on charts B-1 and B-2.

**Legal Liability**
- Level 0: **n/a**
- Level 1: **No**
- Level 2: **Maybe**
- Level 3: **Likely**
- Level 4: **Likely**
- Level 5: **Yes**
- Level 6: **Yes**

### 5.0 Additional Considerations and Future Research

The grid depicted in Figure B-1 and Figure B-2 is a strawman variant, meaning that the indications shown are an initial populating of the grid. Additional research is needed to explore the designations and ascertain whether the initial indications might be advisedly changed or possibly transformed into some other kind of designations, such as numeric scores or weights.

Another aspect of additional research involves the APL/UPL key factors that are utilized in this strawman variant. There are other ways to portray the factors, along with the possibility of adding factors or possibly opting to excise some of the factors from the grid. Research on such modifications is encouraged. As a final point, there are potentially greater questions that arise from the grid, alluded to earlier in the discussion of the prior sections, entailing what actions would be taken if indeed AILR can achieve the autonomous levels of Level 4, Level 5, and Level 6. There remain many such open issues, each deserving of suitable attention.

The FTC observed that the practice-of-law is being buffeted and disrupted by a multitude of societal, economic, and technological changes, as stated in a 2016 memorandum [47]:

"The legal services marketplace has experienced a number of changes in recent years. These trends include: client demands for more cost-effective and efficient services; unbundling of services and disaggregation of legal matters across multiple service providers; development of new billing models and law firm models; geographic expansion of law firms and other legal services providers; provision by non-law firms of certain services previously obtained exclusively from law firms; increased use of automation technologies; online matching, reviewing, and ranking of lawyers; and use of Internet, World Wide Web, and related computer technologies to deliver legal services. In particular, the increased use of computer, software, and online technologies has enabled non-lawyers to provide many services that historically were provided exclusively by lawyers and traditional law firms."

As pointed out in the FTC commentary, legal technologies are increasingly enabling non-lawyers to provide legal services that would normally be considered more so UPL then APL. The next step would seem to be excising the need for a non-lawyer, making use of an autonomous AI Legal Reasoning system in place of any human-based assistance or intervention in delivering legal services and legal advice [51] [53]. That day has not yet arrived [5] [35], but the future appears to encompass such a possibility and it is worthwhile today to examine how the legal profession might need to inexorably adjust in the face of such a significant disruption.

This paper has provided and explored a newly derived instrumental grid depicting the key characteristics underlying APL and UPL as they apply to the AILR autonomous levels and has sought to provide key insights and spur informed discussions regarding the furtherance of crucial practice-of-law deliberations.

### About the Author

Dr. Lance Eliot is the Chief AI Scientist at Techbrium Inc. and a Stanford Fellow at Stanford University in the CodeX: Center for Legal Informatics. He previously was a professor at the University of Southern California (USC) where he headed a multi-disciplinary and pioneering AI research lab. Dr. Eliot is globally recognized for his expertise in AI and is the author of highly ranked AI books and columns.

**Figure A-1**

| \multicolumn{5}{c}{AI & Law: Levels of Autonomy For AI Legal Reasoning (AILR)} | | | | |
|---|---|---|---|---|
| Level | Descriptor | Examples | Automation | Status |
| 0 | No Automation | Manual, paper-based (no automation) | None | De Facto - In Use |
| 1 | Simple Assistance Automation | Word Processing, XLS, online legal docs, etc. | Legal Assist | Widely In Use |
| 2 | Advanced Assistance Automation | Query-style NLP, ML for case prediction, etc. | Legal Assist | Some In Use |
| 3 | Semi-Autonomous Automation | KBS & ML/DL for legal reasoning & analysis, etc. | Legal Assist | Primarily Prototypes & Research Based |
| 4 | AILR Domain Autonomous | Versed only in a specific legal domain | Legal Advisor (law fluent) | None As Yet |
| 5 | AILR Fully Autonomous | Versatile within and across all legal domains | Legal Advisor (law fluent) | None As Yet |
| 6 | AILR Superhuman Autonomous | Exceeds human-based legal reasoning | Supra Legal Advisor | Indeterminate |

*Figure 1: AI & Law - Autonomous Levels by Rows*          *Source Author: Dr. Lance B. Eliot*

V1.3

**Figure A-2**

## AI & Law: Levels of Autonomy For AI Legal Reasoning (AILR)

| | Level 0 | Level 1 | Level 2 | Level 3 | Level 4 | Level 5 | Level 6 |
|---|---|---|---|---|---|---|---|
| **Descriptor** | No Automation | Simple Assistance Automation | Advanced Assistance Automation | Semi-Autonomous Automation | AILR Domain Autonomous | AILR Fully Autonomous | AILR Superhuman Autonomous |
| **Examples** | Manual, paper-based (no automation) | Word Processing, XLS, online legal docs, etc. | Query-style NLP, ML for case prediction, etc. | KBS & ML/DL for legal reasoning & analysis, etc. | Versed only in a specific legal domain | Versatile within and across all legal domains | Exceeds human-based legal reasoning |
| **Automation** | None | Legal Assist | Legal Assist | Legal Assist | Legal Advisor (law fluent) | Legal Advisor (law fluent) | Supra Legal Advisor |
| **Status** | De Facto – In Use | Widely In Use | Some In Use | Primarily Prototypes & Research-based | None As Yet | None As Yet | Indeterminate |

*Figure 2: AI & Law - Autonomous Levels by Columns*                    *Source Author: Dr. Lance B. Eliot*

V1.3

**Figure B-1**

## The Practice of Law and Levels of Autonomy For AI Legal Reasoning (AILR)

| Level | Descriptor | Provides Legal Advice | Asserts Practices Law | Lawyer-Client Relationship | Qualified in Law | Incurs Duty of Care | Legal Confidentiality | Enforceable Prof Conduct | Malpractice Susceptible | Legal Liability |
|-------|------------|----------------------|----------------------|---------------------------|-----------------|--------------------|----------------------|-------------------------|------------------------|-----------------|
| 0 | No Automation | n/a | n/a | n/a | n/a | n/a | n/a | n/a | n/a | n/a |
| 1 | Simple Assistance Automation | No | No | No | No | No | No | No | No | No |
| 2 | Advanced Assistance Automation | Maybe | No | No | No | No | No | No | No | Maybe |
| 3 | Semi-Autonomous Automation | Yes | No | No | Minimal | No | No | No | No | Likely |
| 4 | AILR Domain Autonomous | Yes | Yes | Partial | Partial | Likely | Likely | Likely | Likely | Likely |
| 5 | AILR Fully Autonomous | Yes | Yes | Yes | Yes | Yes | Yes | Yes | Yes | Yes |
| 6 | AILR Superhuman Autonomous | Yes Plus | Yes Plus | Yes | Yes Plus | Yes | Yes | Yes | Yes | Yes |

*Figure 2: AI & Law - The Practice of Law and LoA AILR by Rows*  Strawman Variant  *Source Author: Dr. Lance B. Eliot*  V1.3

**Figure B-2**

## The Practice of Law and Autonomous Levels of AI Legal Reasoning (AILR)

| | Level 0 | Level 1 | Level 2 | Level 3 | Level 4 | Level 5 | Level 6 |
|---|---|---|---|---|---|---|---|
| Descriptor | No Automation | Simple Assistance Automation | Advanced Assistance Automation | Semi-Autonomous Automation | AILR Domain Autonomous | AILR Fully Autonomous | AILR Superhuman Autonomous |
| **Provides Legal Advice** | n/a | No | Maybe | Yes | Yes | Yes | Yes Plus |
| **Asserts Practices Law** | n/a | No | No | No | Yes | Yes | Yes Plus |
| **Lawyer-Client Relationship** | n/a | No | No | No | Partial | Yes | Yes |
| **Qualified in Law** | n/a | No | No | Minimal | Partial | Yes | Yes Plus |
| **Incurs Duty of Care** | n/a | No | No | No | Likely | Yes | Yes |
| **Legal Confidentiality** | n/a | No | No | No | Likely | Yes | Yes |
| **Enforceable Prof Conduct** | n/a | No | No | No | Likely | Yes | Yes |
| **Malpractice Susceptible** | n/a | No | No | No | Likely | Yes | Yes |
| **Legal Liability** | n/a | No | Maybe | Likely | Likely | Yes | Yes |

*Strawman Variant*

*Figure 1: AI & Law – The Practice of Law and LoA AILR by Columns*          Source Author: Dr. Lance B. Eliot          V1.3

**Figure B-3**

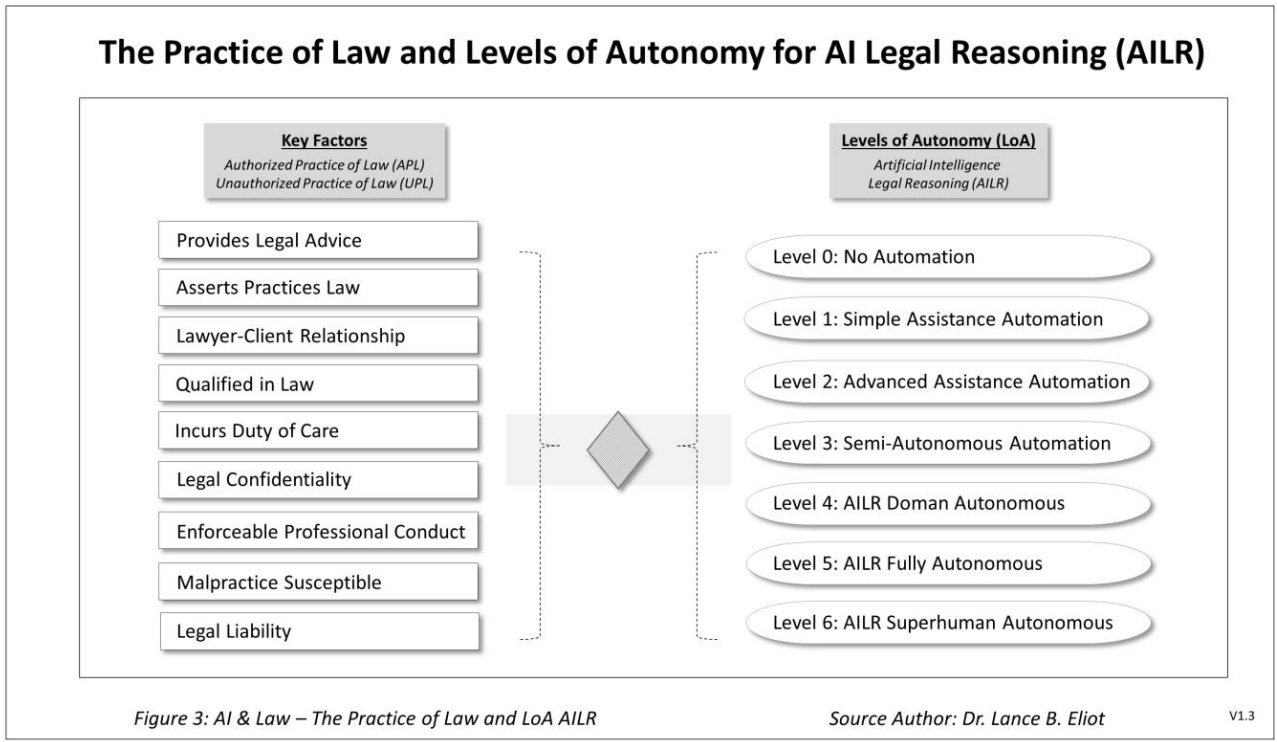

The Practice of Law and Levels of Autonomy for AI Legal Reasoning (AILR)

**Key Factors**
*Authorized Practice of Law (APL)*
*Unauthorized Practice of Law (UPL)*

**Levels of Autonomy (LoA)**
*Artificial Intelligence*
*Legal Reasoning (AILR)*

Provides Legal Advice

Asserts Practices Law

Lawyer-Client Relationship

Qualified in Law

Incurs Duty of Care

Legal Confidentiality

Enforceable Professional Conduct

Malpractice Susceptible

Legal Liability

Level 0: No Automation

Level 1: Simple Assistance Automation

Level 2: Advanced Assistance Automation

Level 3: Semi-Autonomous Automation

Level 4: AILR Doman Autonomous

Level 5: AILR Fully Autonomous

Level 6: AILR Superhuman Autonomous

*Figure 3: AI & Law – The Practice of Law and LoA AILR*          *Source Author: Dr. Lance B. Eliot*          V1.3